\documentclass[aps,twocolumn,noshowpacs,nobalancelastpage,tightenlines,floatfix]{revtex4}
\usepackage{amsmath}
\usepackage{amssymb}
\usepackage{graphicx}
\usepackage{dcolumn}
\usepackage{bm}

\begin{document}

\title{Temporal Dynamics of Photon Pairs Generated by an Atomic Ensemble}
\author{S. V. Polyakov, C.~W. Chou, D. Felinto and H.~J.~Kimble}
\affiliation{Norman Bridge Laboratory of Physics 12-33\\
California Institute of Technology, Pasadena, CA 91125}

\begin{abstract}
The time dependence of nonclassical correlations is investigated for two
fields $(1,2)$ generated by an ensemble of cold Cesium atoms via the
protocol of Duan et al. [Nature \textbf{414}, 413 (2001)]. The correlation
function $R(t_{1},t_{2})$ for the ratio of cross to auto-correlations for
the $(1,2)$ fields at times $(t_{1},$ $t_{2})$ is found to have a maximum
value $R^{\max }=292\pm 57$, which significantly violates the Cauchy-Schwarz
inequality $R\leq 1$ for classical fields. Decoherence of quantum
correlations is observed over $\tau _{d}\simeq 175$ ns, and is described by
our model, as is a new scheme to mitigate this effect.
\end{abstract}

\pacs{PACS Numbers}
\maketitle

In recent years quantum measurement combined with conditional quantum
evolution has emerged as a powerful paradigm for accomplishing diverse tasks
in quantum information science \cite{bennett93,raussendorf01,knill01,duan01}%
. For example, Duan, Lukin, Cirac and Zoller (\textit{DLCZ}) \cite{duan01}
have proposed a scheme for the realization of scalable quantum communication
networks that relies upon entanglement created probabilistically between
remotely located atomic ensembles. By utilizing successful measurements to
condition subsequent steps in their protocol, \textit{DLCZ} have developed a
scheme that has built-in quantum memory, entanglement purification and
resilience to realistic sources of noise, thereby enabling a quantum
repeater architecture to overcome photon attenuation \cite{c16,c17}.

Central to the \textit{DLCZ} protocol is the ability to \textit{write} and
\textit{read} collective spin excitations \textit{into} and \textit{out of}
an atomic ensemble, with efficient conversion of discrete spin excitations
to single-photon wavepackets. Observations of the resulting non-classical
correlations between the optical fields generated from \textit{writing} and
\textit{reading} such spin excitations have recently been reported by
several groups, both at the single-photon level \cite%
{kuzmich03,chou04,jiang03} as appropriate to the protocol of \textit{DLCZ}
and in a regime of large photon number $n\sim 10^{3}-10^{7}$ \cite%
{vanderwal03}. Generation and detection efficiencies have now been improved
so that excitation stored within an atomic ensemble can be employed as a
controllable source for single photons \cite{chou04}.

A critical aspect of such single-photon wavepackets is that they are emitted
into well-defined spatio-temporal modes to enable quantum interference
between emissions from separate ensembles (e.g., for entanglement based
quantum cryptography \cite{duan01}). However, with the exception of the
verification of the time-delay implicit for the Raman processes employed
\cite{vanderwal03}, experiments to date have investigated neither the time
or spatial dependence of quantum correlations for the emitted fields from
the atomic ensemble. The high efficiencies achieved in Ref. \cite{chou04}
now enable such an investigation for the temporal properties of nonclassical
correlations between emitted photon pairs, which we report in this Letter.

\begin{figure}[tb]
\includegraphics[width=8.6cm]{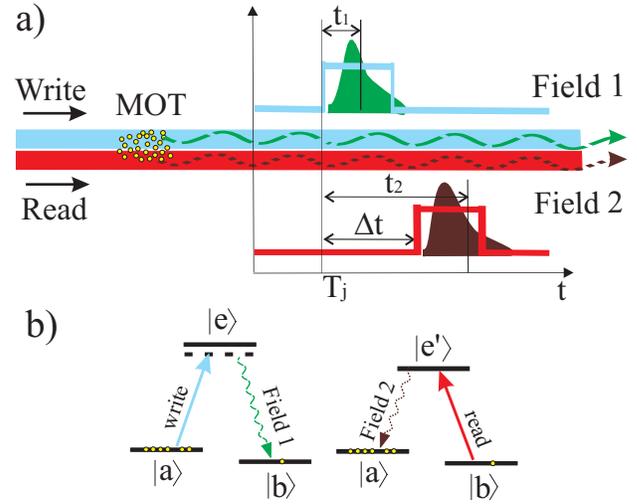}
\caption{(a) Schematic of experiment. \textit{Write} and \textit{read}
pulses propagate into a cloud of cold Cs atoms (MOT) at times $T_{j}$ and $%
T_{j}+\Delta t$ respectively, and generate the correlated output fields $1$
and $2$. Quantum correlations for these fields at times $(t_{1},t_{2})$ are
investigated by way of photoelectric detection. (b) The relevant atomic
level scheme.}
\label{schematic}
\end{figure}

Specifically, we study the time dependence of quantum correlations for
photons emitted from an ensemble of cold Cesium atoms, with photon pairs
created sequentially by classically controlled \textit{write} and \textit{%
read} pulses. Large violations of the Cauchy-Schwartz inequality $R\leq 1$
for the ratio of cross to auto-correlations are observed for the generated
fields, with $R^{\max }=292\pm 57\nleq 1$. By contrast, previous
measurements have reported violations $R\nleq 1$ only for detection events
integrated over the entire durations of the \textit{write} and \textit{read}
pulses ($R=1.84\pm 0.06$ in Ref. \cite{kuzmich03}, $R=1.34\pm 0.05$ in Ref.
\cite{jiang03}, and $R=53\pm 2$ in Ref. \cite{chou04}). We also map the
decay of quantum correlations by varying the time delay between the \textit{%
write} and \textit{read}\ pulses, and find a decoherence time $\tau
_{d}\simeq 175$ ns. We have developed a model to describe the decoherence
and find good correspondence with our measurements. This model is utilized
to analyze a new proposal that should extend the correlation times to beyond
$10$ $\mu $s, which would allow for entanglement between atomic ensembles on
the scale of several kilometers.

Our experimental procedure is illustrated in Figure \ref{schematic}. An
initial \textit{write} pulse at $852$ nm creates a state of collective
excitation in an ensemble of cold Cs atoms, as heralded by a photoelectric
detection event from the Raman field $1$. After a user-programmable delay $%
\Delta t$, a \textit{read} pulse at $894$ nm converts this atomic excitation
into a field excitation of the Raman field $2$. The (\textit{write}, \textit{%
read}) pulses are approximate coherent states with mean photon numbers ($%
10^{3}$,$10^{5}$), respectively, and are focussed into the MOT\ as TEM$_{00}$
Gaussian beams with orthogonal polarizations and beam waist $w_{0}=30$ $\mu $%
m. This scheme is implemented in an optically thick sample of four-level
atoms, cooled and trapped in a magneto-optical trap (MOT) \cite{metcalf99}.
In particular, we utilize the ground hyperfine levels $6S_{1/2},F=\{4;3\}$
of atomic Cs (labelled $\{|a\rangle ;|b\rangle \}$), and excited levels $%
\{6P_{3/2},F=4;6P_{1/2},F=4\}$ of the $D_{2},D_{1}$ lines at $\{852;894\}$
nm (labelled $\{|e\rangle ;|e^{\prime }\rangle \}$).

Each attempt to generate a correlated pair of photons in the $(1,2)$ fields
is preceded by shutting off the trapping light for $700$ ns. The re-pumping
light is left on for an additional $300$ ns in order to empty the $|b\rangle
$ state, thus preparing the atoms in $|a\rangle $. The $j^{\text{th}}$ trial
of a protocol is initiated at time $T_{j}$ when a rectangular pulse from the
\textit{write} laser beam, $150$ ns in duration (FWHM) and tuned $10$ MHz
below the $|a\rangle \rightarrow |e\rangle $ transition, induces spontaneous
Raman scattering to level $|b\rangle $ via $|a\rangle \rightarrow |e\rangle
\rightarrow |b\rangle $. The \textit{write} pulse is sufficiently weak so
that the probability to scatter one Raman photon into a forward propagating
wavepacket is much less than unity for each pulse. Detection of one photon
from field $1$ results in a \textquotedblleft spin\textquotedblright\
excitation to level $|b\rangle $, with this excitation distributed in a
symmetrized, coherent manner throughout the sample of $N$ atoms illuminated
by the \textit{write} beam \cite{duan01}. Regardless of successful detection
of a photon in field $1$, we next address the atomic ensemble with a \textit{%
read} pulse at a time $T_{j}+\Delta t$, where $\Delta t$ is controlled by
the user. The \textit{read} light is a rectangular pulse, $120$ ns in
duration, tuned to resonance with the $|b\rangle \rightarrow |e^{\prime
}\rangle $ transition.

To investigate the photon statistics, we use four avalanche photodetectors,
a pair for each field $i$, labelled as $D_{iA,iB}$, which are activated at $%
(T_{j},T_{j}+\Delta t)$ with $i=(1,2)$, respectively, for $200$ ns
for all experiments. The quantity $p_{\tau }(t_{l},t_{m})$ is
defined as the joint
probability for photoelectric detection from field $l$ in the interval $%
[T_{j}+t_{l},T_{j}+t_{l}+\tau ]$ and for an event from field $m$ in the
interval $[T_{j}+t_{m},T_{j}+t_{m}+\tau ]$, where $l$ and $m$ equal $1$ or $%
2 $. $p_{\tau }(t_{l},t_{m})$ is determined from the record of time-stamped
detection events at $D_{1A,1B},D_{2A,2B}$. In a similar fashion, $q_{\tau
}(t_{l},t_{m})$ gives the joint probability for detection for fields $(l,m)$
in the intervals $([T_{j}+t_{l},T_{j}+t_{l}+\tau
],[T_{k}+t_{m},T_{k}+t_{m}+\tau ])$ for two trials $k\neq j$.

Following Refs. \cite{kuzmich03,chou04}, we introduce the time-dependent
ratio $R_{\tau }(t_{1},t_{2})$ of cross-correlation to auto-correlation for
the $(1,2)$ fields, where%
\begin{equation}
R_{\tau }(t_{1},t_{2})\equiv \frac{\lbrack p_{\tau }(t_{1},t_{2})]^{2}}{%
p_{\tau }(t_{1},t_{1})p_{\tau }(t_{2},t_{2})}\text{ .}  \label{R}
\end{equation}%
This ratio is constrained by the inequality $R_{\tau }(t_{1},t_{2})\leqslant
1$ for all fields for which the Glauber-Sudarshan phase-space function is
well-behaved (i.e., \textit{classical} fields) \cite{kuzmich03,clauser74}.
Beyond enabling a characterization of the quantum character of the $(1,2)$
fields in a model independent fashion, measurements of $R_{\tau
}(t_{1},t_{2})$ also allow inferences of the quantum state for collective
excitations of single spins within the atomic ensemble.

\begin{figure}[tb]
\includegraphics[width=8.6cm]{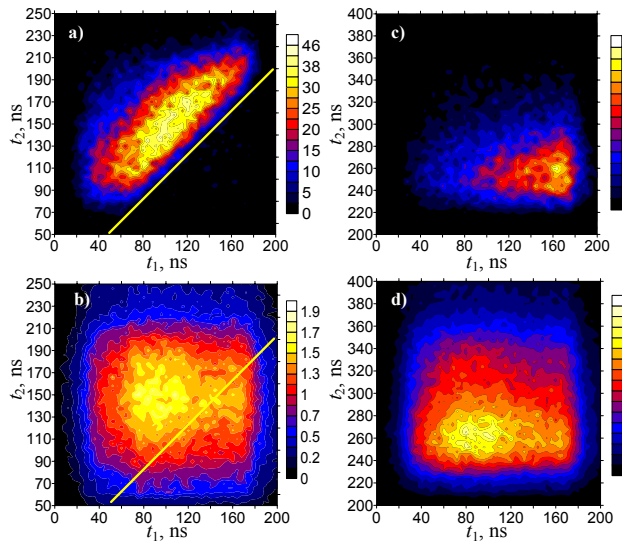}
\caption{Probability for joint detection from the fields $(1,2)$ at times $%
(t_{1},t_{2})$ with origin at the beginning of the \textit{write} pulse. (a)
$p_{\protect\tau }(t_{1},t_{2})$, and (b) $q_{\protect\tau }(t_{1},t_{2})$
for overlapped \textit{write} and \textit{read} pulses, $\Delta t=50$ ns,
with the solid line corresponding to $t_{2}=t_{1}$. (c) $p_{\protect\tau %
}(t_{1},t_{2})$ and (d) $q_{\protect\tau }(t_{1},t_{2})$ for consecutive
\textit{write} and \textit{read} pulses, $\Delta t=200$ ns. In all cases,
the bin size $\protect\tau =4$ ns, and the joint probabilities $p_{\protect%
\tau }$ and $q_{\protect\tau }$ have been scaled by $10^{9}$.}
\label{corr3d}
\end{figure}

The first step in the determination of $R_{\tau }(t_{1},t_{2})$ is the
measurement of the joint probability $p_{\tau }(t_{1},t_{2})$ for the $(1,2)$
fields, and for comparison, $q_{\tau }(t_{1},t_{2})$ for independent trials.
In our experiment, we focus on two cases: (\textit{I}) nearly simultaneous
application of \textit{write} and \textit{read} pulses with offset $\Delta
t=50$ ns less than the duration of either pulse, and (\textit{II})
consecutive application of \textit{write} and \textit{read} pulses with $%
\Delta t=200$ ns longer than the \textit{write},\textit{\ read} durations.
Results for $p_{\tau }(t_{1},t_{2})$ and $q_{\tau }(t_{1},t_{2})$ are
presented in Fig. \ref{corr3d} as functions of the detection times $%
(t_{1},t_{2})$ for the fields $(1,2)$. For both $\Delta t=50$ and $200$ ns, $%
p_{\tau }(t_{1},t_{2})\gg q_{\tau }(t_{1},t_{2})$, indicating the strong
correlation between fields $1$ and $2$, with the maximal ratio $%
g_{1,2}^{\tau }(t_{1},t_{2})=p_{\tau }(t_{1},t_{2})/q_{\tau
}(t_{1},t_{2})\gtrsim 30$, which is much greater than reported previously
\cite{kuzmich03,jiang03,chou04}. In Fig. \ref{corr3d}, $\tau =4$ ns, leading
to statistical errors of about $8\%$ for the largest values shown.

\begin{figure}[tb]
\includegraphics[width=8.6cm]{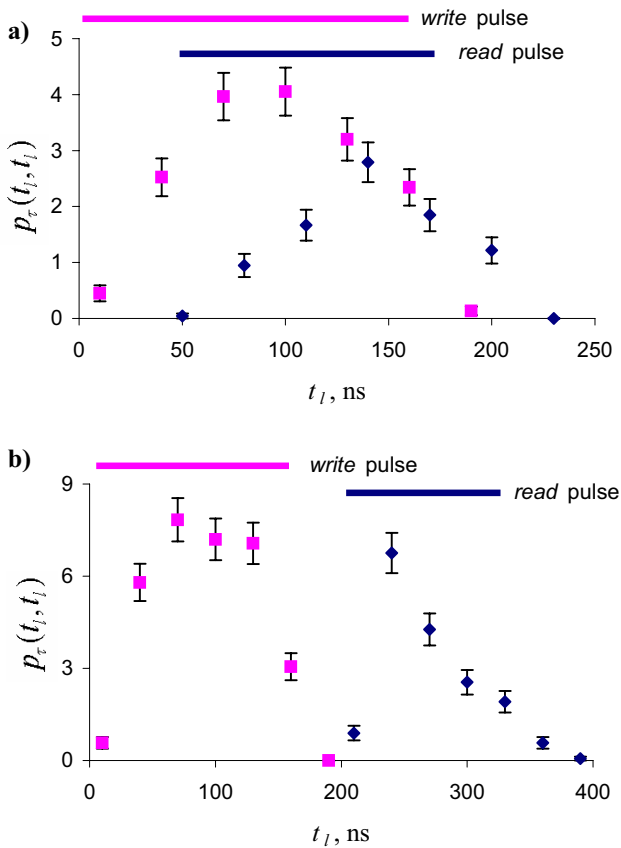}
\caption{Probability of joint detection $p_{\protect\tau }(t_{1},t_{1})$ for
field $1$ (squares) and $p_{\protect\tau }(t_{2},t_{2})$ for field $2$
(diamonds) as functions of respective detection times $t_{1}$ and $t_{2}$.
Bin size $\protect\tau=30$ ns. (a) Overlapping \textit{write} and \textit{%
read} pulses, $\Delta t=50$ ns; (b) Consecutive \textit{write} and \textit{%
read} pulses, $\Delta t=200$ ns.}
\label{corr2d}
\end{figure}

\begin{figure}[tb]
\includegraphics[width=8.6cm]{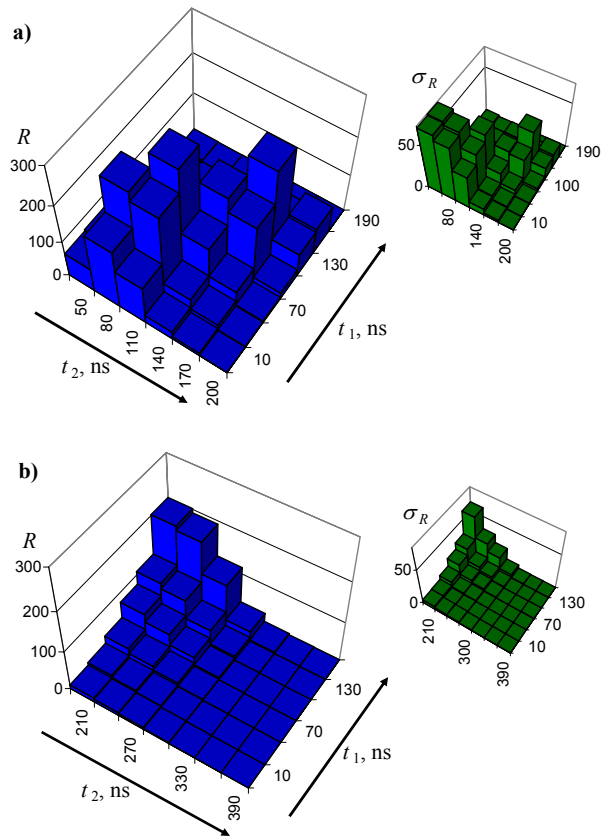}
\caption{The experimentally derived ratio $R_{\protect\tau }(t_{1},t_{2})$
[Eq. \protect\ref{R}] as a function of detection times $(t_{1},t_{2})$ for
the $(1,2)$ fields, with $R_{\protect\tau }\leq 1$ for classical fields. The
left column gives $R_{\protect\tau }(t_{1},t_{2})$ for $\Delta t=50$ ns
(top) and $\Delta t=200$ ns (bottom), while the right column gives the
associated statistical uncertainties. Bin size $\protect\tau =30$ ns.}
\label{Rt}
\end{figure}

In case (\textit{I})\ for nearly simultaneous irradiation with \textit{write}
and \textit{read} pulses, Fig. \ref{corr3d}(a) shows that $p_{\tau
}(t_{1},t_{2})$ peaks along the line $t_{2}-t_{1}=\delta t_{12}\simeq 50$ ns
with a width $\Delta t_{12}\simeq 60$ ns, in correspondence to the delay $%
\delta t_{12}$ and duration $\Delta t_{12}$ for read-out associated with the
transition $|b\rangle \rightarrow |e^{\prime }\rangle \rightarrow |a\rangle $
given an initial transition $|a\rangle \rightarrow |e\rangle \rightarrow
|b\rangle $ \cite{vanderwal03}. Apparently, the qualitative features of $%
p_{\tau }(t_{1},t_{2})$ depend only upon the time difference between photon
detections in fields $1$ and $2$ (i.e., $p_{\tau }(t_{1},t_{2})\approx
F(t_{2}-t_{1})$). In case (\textit{II})\ with the \textit{read} pulse
launched $200$ ns after the \textit{write} pulse, excitation is
\textquotedblleft stored\textquotedblright\ in the atomic ensemble until the
readout. The production of correlated photon pairs should now be distributed
along $t_{2}\simeq \delta t_{12}$ with width $\simeq $ $\Delta t_{12}$.
Instead, as shown in Fig. \ref{corr3d}(c), $p_{\tau }(t_{1},t_{2})$ peaks
towards the end of the \textit{write} pulse (i.e., $t_{1}\gtrsim 100$ ns),
and near the beginning of the \textit{read} pulse (i.e., $200\lesssim
t_{2}\lesssim 300$ ns). Early events for field $1$ lead to fewer correlated
events for field $2$, as $p_{\tau }(t_{1},t_{2})$ decays rapidly beyond the
line $t_{2}-t_{1}=\tau _{d}\simeq 175$ ns. The marked contrast between $%
p_{\tau }(t_{1},t_{2})$ for $\Delta t=50$ and $200$ ns results in a
diminished ability for the conditional generation of single photons from
excitation stored within the atomic ensemble \cite{chou04} and, more
generally, for the implementation of the \textit{DLCZ} protocol for
increasing $\Delta t$. The underlying mechanism is decoherence within the
ensemble, as will be discussed.

Fig. \ref{corr3d}(b, d) displays $q_{\tau }(t_{1},t_{2})$ for independent
trials $j\neq k$. $q_{\tau }(t_{1},t_{2})$ is expected to be proportional to
the product of intensities of the fields $1$ and $2 $, in reasonable
correspondence to the form shown in Fig. \ref{corr3d}(b, d) for our roughly
rectangular \textit{write}, \textit{read} pulses, but distinctively
different from $p_{\tau }(t_{1},t_{2})$ in (a, c).

To deduce $R_{\tau }(t_{1},t_{2})$ from Eq. \ref{R}, we next determine the
joint detection probabilities $p_{\tau }(t_{1},t_{1})$ for field $1$\ and $%
p_{\tau }(t_{2},t_{2})$ for field $2$ from the same record of photoelectric
events as for Fig. \ref{corr3d} (a, c). Since the rate of coincidences for
auto-correlations is roughly $10^{2}$ times smaller than for
cross-correlations for the $(1,2)$ fields, we increase the bin size $\tau $
to $30$ ns to accumulate enough events to reduce the statistical errors to
acceptable levels. Fig. \ref{corr2d} shows the resulting time dependencies
of $p_{\tau }(t_{1},t_{1})$ and $p_{\tau }(t_{2},t_{2})$ for cases (\textit{I%
}, \textit{II}). While the shape of $p_{\tau }(t_{1},t_{1})$ associated with
the \textit{write} pulse does not change with $\Delta t$, the profile of $%
p_{\tau }(t_{2},t_{2})$ from the \textit{read} pulse is affected and
exhibits a rise time that is $\sim 3$ times shorter for $\Delta t=200$ ns
than for $\Delta t=50$ ns. This prompt rise in (b) is consistent with the
observation that stored excitation is efficiently addressed at the beginning
of the \textit{read} pulse for non-overlapping \textit{write}, \textit{read}
pulses, while the longer rise time in (a) results from continuous excitation
and retrieval of atoms from the state $|b\rangle $ for the overlapping case.

We employ the data in Figs. \ref{corr3d}, \ref{corr2d} together with Eq. \ref%
{R} to construct the ratio $R_{\tau }(t_{1},t_{2})$, with the result
presented in Fig. \ref{Rt} \cite{rebin}. Not unexpectedly, the trends for\ $%
R_{\tau }(t_{1},t_{2})$ closely resemble those of the joint probability $%
p_{\tau }(t_{1},t_{2})$ for correlated pair generation previously discussed.
As for the violation of the Cauchy-Schwarz inequality $R_{\tau
}(t_{1},t_{2})\leq 1$ for classical fields \cite{kuzmich03,clauser74}, we
observe maximal violations with $R_{\tau }^{\max }=292\pm 57$ for $\Delta
t=50$ ns and $R_{\tau }^{\max }=202\pm 60$ for $\Delta t=200$ ns ($R_{\tau
}=198\pm 33$ in the neighboring bin). The relatively large errors in $%
R_{\tau }(t_{1},t_{2})$ arise predominantly from the uncertainties in $%
p_{\tau }(t_{1},t_{1})$ and $p_{\tau }(t_{2},t_{2})$ (Fig. \ref{corr2d})
\cite{verify}.

\begin{figure}[tb]
\includegraphics[width=8.6cm]{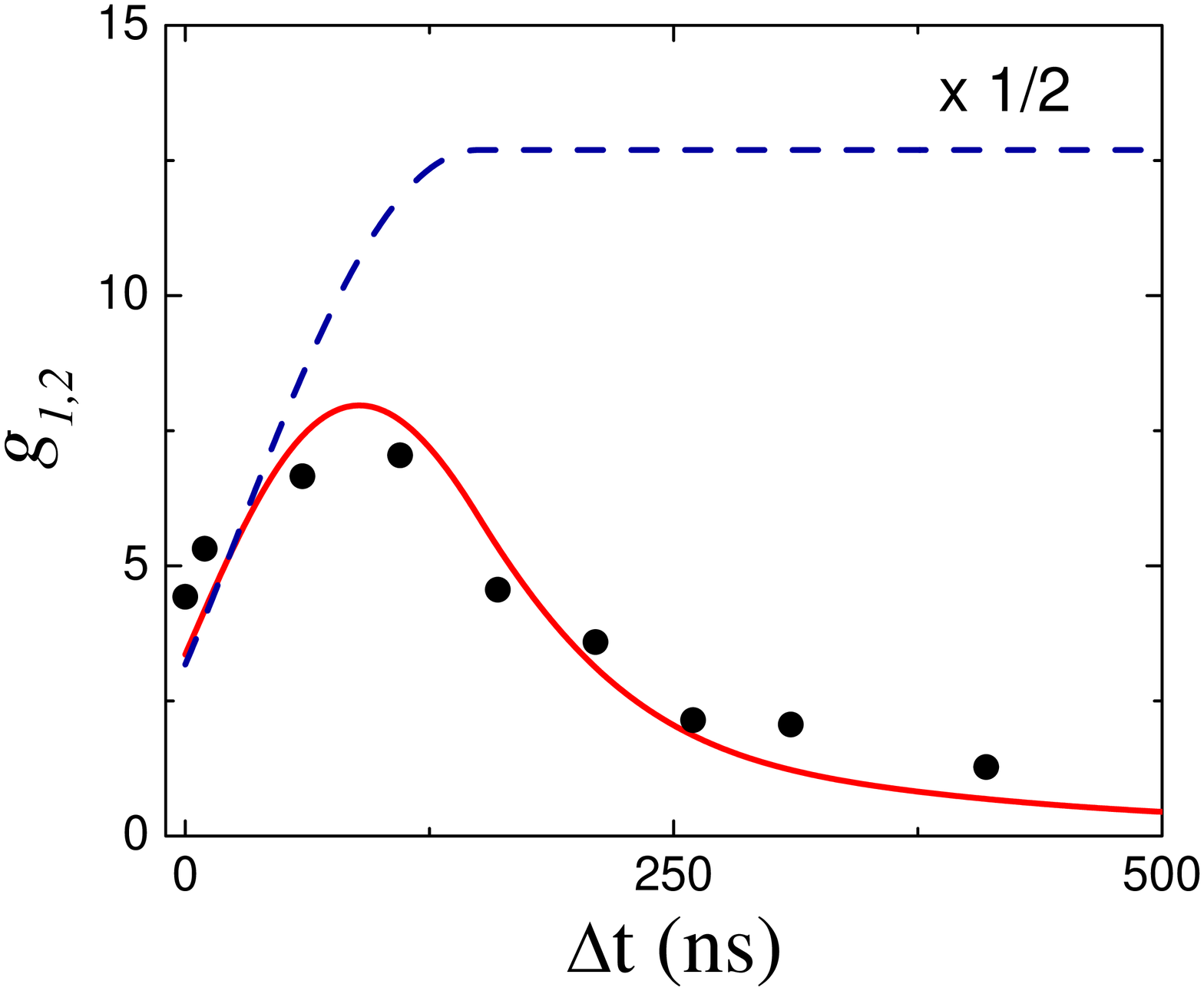}
\caption{Coherence time assessment. Experimentally acquired $g_{1,2}$ (black
dots), theoretical description of the current experiment with $K=1.1$ MHz
(solid line), and the theoretical prediction for a spin polarized $m_{F}=0$
MOT (dotted line).}
\label{decoherence}
\end{figure}

The forms for $p_{\tau }(t_{1},t_{2})$ and $R_{\tau }(t_{1},t_{2})$ for the
cases $\Delta t=50$ and $200$ ns imply a decoherence process operative on a
time scale $\tau _{d}\sim 175$ ns. To investigate this decay, we have
performed a separate set of experiments with the delay $\Delta t$ varied $%
0\leq \Delta t\leq 400$ ns. For each $\Delta t$ we determine the normalized
correlation function $g_{1,2}^{\tau }$ from the ratio of integrated
coincidence counts to singles counts over the entire detection window (i.e.,
$\tau =200$ ns), with the results presented in Fig. \ref{decoherence}.

In Fig. \ref{decoherence}, the initial growth of $g_{1,2}^{\tau }$ for small
$\Delta t$ is due to the finite time required to produce sequentially
photons in the $(1,2)$ fields, which is already evident in Fig. \ref{corr3d}%
. More troublesome is the rapid decay of $g_{1,2}^{\tau }$. A likely
candidate responsible for this decay is Larmor precession among the various
Zeeman states of the $F=3,4$ hyperfine levels of the $6S_{1/2}$ ground level.

To investigate this possibility, we have extended the treatment of Ref. \cite%
{duan02} to include the process involving the \textit{read} beam as well as
the full set of Zeeman states for the $F=3,4$ hyperfine levels. The sample
of Cs atoms is assumed to be initially unpolarized and distributed over the
same range of magnetic fields as for the MOT. With \textit{write} and
\textit{read }pulses that approximate those used in our experiment, we
obtain an expression for the probability $\tilde{p}_{\tau }(t_{1},t_{2})$ to
generate a pair of photons at times $(t_{1},t_{2})$ for fields $1$ and $2$
as a function of the offset $\Delta t$. By summing contributions for all $%
(t_{1},t_{2})$ over the detection windows, we arrive at the joint
probability $\tilde{p}_{1,2}(\Delta t)$ that we compare to the measured $%
g_{1,2}(\Delta t)$ by way of a single overall scaling parameter for all $%
\Delta t$, as the rate of single counts in fields $(1,2)$ is measured not to
depend on $\Delta t$ (to within $20\%$) The result is the solid curve in
Fig. \ref{decoherence} that evidently adequately describes the impact of
Larmor precession on our experiment. The form of $\tilde{p}_{1,2}(\Delta t)$
strongly depends upon the inhomogeneity of Zeeman splitting across the MOT,
which is described by the parameter $K=\mu _{B}g_{F_{g}}Lb/h$, where $L$ is
the MOT\ diameter, $b$ is the gradient of the magnetic field for the MOT,
and $g_{F_{g}}$ is the Land\'{e} factor. The curve in Fig. \ref{decoherence}
is the theoretical result for an initially unpolarized sample with $K=1.1$
MHz as for our experiment (i.e., $L\approx 3.6$ mm and $b\approx 8.4$ G/cm).

An obvious remedy for this dephasing is to eliminate the magnetic field
altogether, as by transferring the sample to a dipole-force trap \cite%
{metcalf99}. Alternatively, we have developed a scheme that should allow for
long coherence times even in the presence of the quadrupole field of the MOT
by utilizing only magnetic-field insensitive states. The \textit{write},
\textit{read} beams are polarized $\sigma _{\pm }$ and are aligned along the
$z-$axis of the MOT, which provides the quantization axis. Atoms within the
approximately cylindrical volume illuminated by these beams are initially
spin polarized into $F=3,m_{F}=0$ \cite{cylinder}. The $(1,2)$ fields are
selected to be $\sigma _{\pm }$, which results in spin excitation stored in $%
F=4,m_{F}=0$. The prediction of our model for this new protocol for the same
experimental conditions as before but now with an initially spin polarized
sample is shown as the dashed curve in Figure \ref{decoherence}, resulting
in an increase of more than $3\times $ in $g_{1,2}^{\tau }$, and
significantly extending the decoherence time to more than $\tau _{d}\sim
10\mu $s.

In conclusion, we have reported the first observations of the temporal
dependence of the joint probability $p_{\tau }(t_{1},t_{2})$ for the
generation of correlated photon pairs from an atomic ensemble, which is
critical for the protocol of Ref. \cite{duan01}. Our measurements of $%
p_{\tau }(t_{1},t_{2})$ are an initial attempt to determine the structure of
the underlying two-photon wavepacket \cite{gheri98}. The nonclassical
character of the emitted $(1,2)$ fields has been tracked by way of time
dependence of the ratio $R_{\tau }(t_{1},t_{2})$, with $R^{\max }=292\pm
57\nleq 1$. Decoherence due to Larmor precession has been characterized and
identified as a principal limitation of the current experiment. A new scheme
for effectively eliminating this decay process has been proposed and
analyzed, and could be important for the experimental realization of
scalable quantum networks \cite{duan01} as well as for an improved source
for single photons \cite{chou04}.

This work is supported by ARDA, the Caltech MURI Center for Quantum Networks, and the NSF.


\begin{thebibliography}{99}
\bibitem{rebin} Before applying Eq. \ref{R}, we reprocess the data for $%
p_{\tau }(t_{1},t_{2})$ in Fig. \ref{corr3d} for $\tau =30$ ns bin size.
\bibitem{verify} To verify the validity of our experimental procedures, we
calculate $R(t_{1},t_{2})$ for detection events for the $(1,2)$ fields from
different trials $j\neq k$, for which $R(t_{1}^{j},t_{2}^{k})$ is expected
to equal unity. This result is confirmed experimentally to within a
statistical uncertainty of less than $4\%$.
\bibitem{cylinder} Note that the beam waist $w_{0}=30\mu $m for the \textit{%
write}, \textit{read} fields is much smaller than the characteristic
dimension $l>1$mm of the MOT.
\end{thebibliography}
\end{document}